\documentclass[12pt]{iopart}
\usepackage[colorlinks=true,citecolor=Cerulean,linkcolor=RubineRed,urlcolor=Cerulean]{hyperref}
\hypersetup{breaklinks=true}
\usepackage{graphicx}
\usepackage{color}
\usepackage[usenames,dvipsnames]{xcolor}
\usepackage{epstopdf}
\usepackage{units}
\usepackage{subcaption}
\newcommand{\Heff}{H_\mathrm{eff}}
\newcommand{\im}{\mathrm{i}}

\begin{document}

\title[Expansion of a one-dimensional Bose gas]
{Expansion of a one-dimensional Bose gas: the role
of interactions and kinetic-energy driving}

\author{E.B.~Molinero, C.E.~Creffield and F.~Sols}
\address{Departamento de F\'isica de Materiales, Universidad
Complutense de Madrid, E-28040 Madrid, Spain}
\ead{c.creffield@fis.ucm.es}

\date{\today}

\begin{abstract}
We study the expansion of a one-dimensional boson gas by 
initialising it in a small region of a chain, and then suddenly allowing
it to expand into the remainder of the chain.
We consider three initial ground-state configurations: the Mott insulator, the conventional superfluid, whose momentum density is sharply peaked at zero 
momentum, and the cat-like state with momentum peaks at $\pm \pi/2$, produced by kinetic driving, the latter being a particular case of a flat-band system. In turn, we consider three types of expansion: spectroscopic (with interactions tuned to zero), dynamic (with standard short-range repulsive interactions), and under kinetic driving. The numerical calculations are exact. We compute the momentum and real-space one-particle densities, as well as the two-particle momentum correlations. We find that the spectroscopic time-of-flight experiment reflects the initial momentum distribution except for the larger number of momentum states and at high momenta. For the dynamic expansion starting from an insulator, we recover the non-equilibrium quasi-condensation into momenta $\pm \pi/2$, provide a physical explanation in terms of interacting bosons that is confirmed by the numerical simulation, and note the existence of nontrivial correlations in the momentum distribution. Under kinetic driving the expansion is comparatively slow, but we conjecture that at high densities it will be much faster. We compare various measures of the two-particle momentum correlations, noting that some of them tend to conceal the possible cat-like structure of a many-body state.
\end{abstract}


\bigskip
\submitto{\jpb}
\maketitle

\section{Introduction}
The expansion of a quantum gas after release of the trapping potential has been a fundamental experimental tool since the very dawn of the field of degenerate dilute quantum gases, both bosonic  \cite{95AN,95DA,95BR-hulet} and fermionic \cite{98DE-Jin,03GR-jin}. The underlying principle is that when the expanding cloud is much larger than the initially trapped gas, the density distribution provides a map of the velocity distribution before the expansion, provided that interactions do not play an important role during the enlargement of the cloud. It has been recognized that for expansions in two and three dimensions, and if the initial densities are not too large, interactions are unimportant in a time of flight experiment \cite{04PA,08GE,10KU,16CH,18CA,19CA,20TE}. The situation is different in one dimension, where interactions during the expansion do matter unless they are made artificially zero through the use of Feshbach resonances \cite{12SC,13R0,13VI}.
When the interactions are important, the expansion of quantum gases challenges and stimulates our understanding of non-equilibrium many-body physics not only in the continuum \cite{96CA,96KA,06KI}, but also on a lattice \cite{12SC,05RI,11JR}.

When the release of the trapping potential includes the removal of the optical lattice, so that the atomic cloud expands in free space, the emission of atom waves from different sites yields interference patterns that convey information on the initial system \cite{08GE,02GR-1,02GR-2,10FA}. Conversely, another interesting scenario is to remove the superimposed trapping potential but to retain the optical lattice, so that the cloud expands in a discretised space rather than a continuum \cite{12SC,13R0,13VI,11JR}. The theoretical and experimental study of the expansion of a quantum gas within a lattice is interesting in its own right \cite{08TO,14BO}, but sometimes it can be just a tool to perform computationally affordable simulations of the expansion of a quantum gas in free space \cite{10KU}.

As a spectroscopic tool, time-of-flight (TOF) experiments can serve to investigate stationary states of exotic many-body Hamiltonians that result from applying some type of external time-periodic driving \cite{17EC,16CR}, a technique known as ``Floquet engineering''. 
Recently, investigations have been made of the behavior of a one-dimensional boson system, described with the Bose-Hubbard model, whose kinetic energy is made to oscillate with a vanishing time average \cite{18PI,19PI}. At high frequencies, the effective Hamiltonian resulting from this kinetic driving is such that first-order single-particle hopping is suppressed, but hopping processes of even order (including assisted tunneling) are permitted. At small amplitudes the system is a Mott insulator. Remarkably, at  higher amplitudes the system acquires an exotic form of superfluidity, based on 
a fragmented, cat-like condensate whose branches peak at momenta $\pm \pi/2$ in units of the inverse lattice spacing. The main results have been shown to be insensitive to variations in the signal shape and in the switching protocol of the kinetic driving \cite{21MA}.

Another attractive feature of rapid kinetic driving is that it generates a particular instance of a flat-band system, a topic of great current interest due to its realization in twisted bilayer graphene \cite{18CA-Jarillo,20LI} and cold atom systems \cite{10AP,18TO-Paivi,21JU-Paivi}.

In this paper we investigate the possibility of probing this exotic ground state with TOF experiments. We consider three types of expansion. The most informative one is that in which both the driving amplitude and the interaction are made zero right at the start of the expansion. The flight of the fragmented condensate can then be understood quite accurately in terms of the momentum distribution and correlations of the ground state just before the expansion, 
except for high initial momenta, in which case the lattice structure has an effect. We also study the case of an expansion in which the driving is turned off, but a strong onsite interaction remains. Finally, we consider the case in which the kinetic driving remains switched on while the strongly interacting quantum gas expands.

In all cases we compare and contrast our results with those obtained for the conventional (undriven) Bose-Hubbard (BH) model, which has two types of ground state: a Mott insulator and a superfluid formed by a quasi-condensate with its mildly divergent depletion cloud. As expected, the Mott insulator phases of the conventional BH and the kinetically-driven (KD) systems behave almost identically. By contrast the superfluid phase of the KD Bose-Hubbard system, supported by two macroscopic cat branches at nonzero momenta, behaves very differently from the superfluid phase of the conventional BH system centered around zero momentum.

In the case where the expansion takes place with a strong interaction (which we refer to as dynamic expansion), we reproduce features such as the dynamic quasi-condensation into states of momenta  $\pm \pi/2$ \cite{04RI,05RI,15VI}. We find that this strong-interaction evolution is quite robust in the sense of being rather insensitive to the initial state, and provide a physical explanation in terms of interacting bosons that also gives a good account of the dynamical condensation into zero momentum when the condensate wave components meet at the opposite extreme of the ring. We report that similar results are obtained if the expansion takes place within a longer segment delimited by hard walls.

By computing the two-particle momentum correlations, we also find that the dynamic condensation occurs in a cat-like form, with the two macroscopic branches flying apart in opposite directions. 
We show in \ref{app_g2} that such extreme momentum correlations are easily missed if one uses the normalized second-order correlation function $g^{(2)}$ because of the excessive weight it gives to unlikely momentum values. We argue that it is not clear to what extent the large correlations found in the momentum occupations will survive in the thermodynamic limit.

Section \ref{sec_models} of this paper is devoted to a presentation of the driving and interaction models used in this work. Section \ref{sec_spectroscopy} deals with the simulation results for the spectroscopic (with zero interaction) TOF experiments. For the initial Mott insulator, we derive analytical expressions for the two-particle momentum correlation in section \ref{mc-MI}.
In section \ref{sec_dynamic}, we investigate the dynamic case of an expansion with strong interactions. While we confirm previously reported results, we provide new physical insights, as explained in section \ref{mc-MI-dyn}. Section \ref{sec_kd} deals with the expansion in the presence of kinetic driving. We find that it is slow for reasons that were anticipated in Ref. \cite{19PI} but we predict that it will increase at larger densities.
Section \ref{sec_velocity} is devoted to a discussion of the velocity scales involved in the problem. A summary and final conclusions are given in section \ref{sec_conclusions}.

\section{Models \label{sec_models}}

Our reference model is the Bose-Hubbard Hamiltonian on a chain,
\begin{equation}
\label{BH}
H= \sum_{x}\left[\frac{U}{2}n_x(n_x-1)-J\left(a^{\dagger}_{x}a_{x+1} + \mathrm{H.c.}\right)\right],
\end{equation}
where $J,U \geq 0$ are the hopping and repulsion energies, $n_x=a^{\dagger}_{x}a_{x}$, and  $a_{x}$ annihilates a boson at site $x$. Depending on the ratio $U/J$, this model presents two different, well-known quantum phases. If $U/J\ll 1$, the system is in the superfluid phase, with a macroscopic occupation of the lowest-energy one-atom state. For $U/J \gg 1$ the system is in the Mott insulating phase if the average occupation is commensurate \cite{03GI}.

Refs. \cite{18PI,19PI,21MA} investigated the behavior of the system when the hopping energy is made to oscillate with zero time average,
\begin{equation}
\label{KD}
J(t) = J\cos(\omega t) \, .
\end{equation}
At high frequencies, the dynamics resulting from this kinetic-energy driving is ruled by an effective time-independent Hamiltonian. After a unitary transformation and averaging over one period one obtains \cite{18PI,19PI}
\begin{equation}
	\label{Heff}
		\Heff = \frac{U}{2L} \sum_{lmnp}{\cal J}_0\{2\kappa F(k_p,k_n,k_m,k_l)]\} \ 
		 \delta_{k_p+k_n,k_m+k_l}a^{\dagger}_{k_p}a^{\dagger}_{k_n}a_{k_m}a_{k_l}\, ,
\end{equation}
where 
\begin{equation}
		F(k_p,k_n,k_m,k_l) \equiv \cos(k_l)+\cos(k_m) 
		- \cos(k_n)-\cos(k_p) \, ,
\end{equation}
${\cal J}_0(x)$ is the Bessel function of zeroth order, and $\kappa \equiv J/\omega$. Here and in the following, we take $\hbar = 1$ and measure all energies 
(times) in units of $J$ ($J^{-1}$). The allowed crystal momenta are defined by integer numbers; for instance, $k_p = 2\pi p/L$ with $p \in \{1,...,L\}$. The Kronecker delta ensures momentum conservation mod $2\pi$.

If we shift to a position representation through the transformation 
\begin{equation}
	a_{k} = \frac{1}{\sqrt{L}}\sum_{x=1}^{L}e^{\im k x}a_x \, , \label{akax}
\end{equation}
the resulting Hamiltonian reveals nonlocal correlations involving even-order hopping processes \cite{18PI}.

At $\kappa \sim 0.4$ a crossover takes place between two qualitatively different ground states. At small amplitudes, the system behaves like a Mott insulator, due to the vanishing of the average hopping energy.  For large $\kappa$, correlated hopping takes over and the system becomes an exotic superfluid with a cat-like structure involving two distinct branches characterized by the macroscopic occupation of momenta $\pm \pi/2$. The two branches share a {\it reduction cloud} formed by the pair-like occupation of momenta $k$ and $\pi -k$ with  $k \neq \pm \pi/2$ \cite{19PI}.

All our results for the effective Hamiltonian have been duplicated with the exact numerical resolution of the full time-dependent Hamiltonian, obtained by inserting \eref{KD} into \eref{BH}.
The resulting physics has been shown to be robust against variations of the signal shape and the switching protocol of the kinetic driving \cite{21MA}.
The insensitivity to the choice of initial time within the driving period found in Ref. \cite{21MA} strongly suggests that the micromotion (evolution of the system within the driving period) is featureless, as we have checked explicitly. While this is a general property of the high-frequency limit, in this system it is further reinforced by the effects discussed in Ref. \cite{21MA}.

Although our initial motivation is to investigate the possibility of probing the ground state of \eref{Heff} through a time-of-flight experiment, we also perform a similar study of the conventional, undriven BH model \eref{BH}, for the sake of comparison and to improve our understanding of the conventional case.
In both the conventional BH and the kinetic-driving scenarios, we consider the cases in which, before the expansion, the ground state is a Mott insulator or a superfluid. That initial state is obtained from exact diagonalization of the Hamiltonian prior to the flight.

We study three types of expansions: with zero interaction, with a strong interaction, and in the presence of kinetic-energy driving. In all cases, we perform an exact numerical calculation of the evolution.

In general, we consider the case of $N$ bosons in a lattice with $L$ sites, denoted by $(N,L)$. In Refs.  \cite{18PI,19PI,21MA}, the case (8,8) at equilibrium was considered. Here, we study the expansion from (4,4) to $(4,16)$. We shall compute the one-particle position and momentum densities
\begin{equation}
	\label{dens}
	\rho(x,t) = \langle n_x \rangle_t \, , \; \; \rho(k,t) = \langle n_k \rangle_t \, ,
\end{equation}
where $\langle \cdot \rangle_t$ stands for the quantum average at time $t$.

Importantly, we will also compute the two-particle momentum density
\begin{equation}
	\rho(k,k';t) = \langle n_k n_{k'} \rangle_t \, .
	\label{2rpdm}
\end{equation}
The reason for choosing this measure of two-particle correlations, instead of the more common second-order correlation function $g^{(2)}$, is that in its often used normalized version the latter tends to hide cat-like correlations by assigning too much weight to unlikely momentum values. This point is discussed further in \ref{app_g2}.

\section{Time-of-flight spectroscopy\label{sec_spectroscopy}}

\begin{figure}
\begin{center}
        \includegraphics[width=.6\textwidth,clip=true]{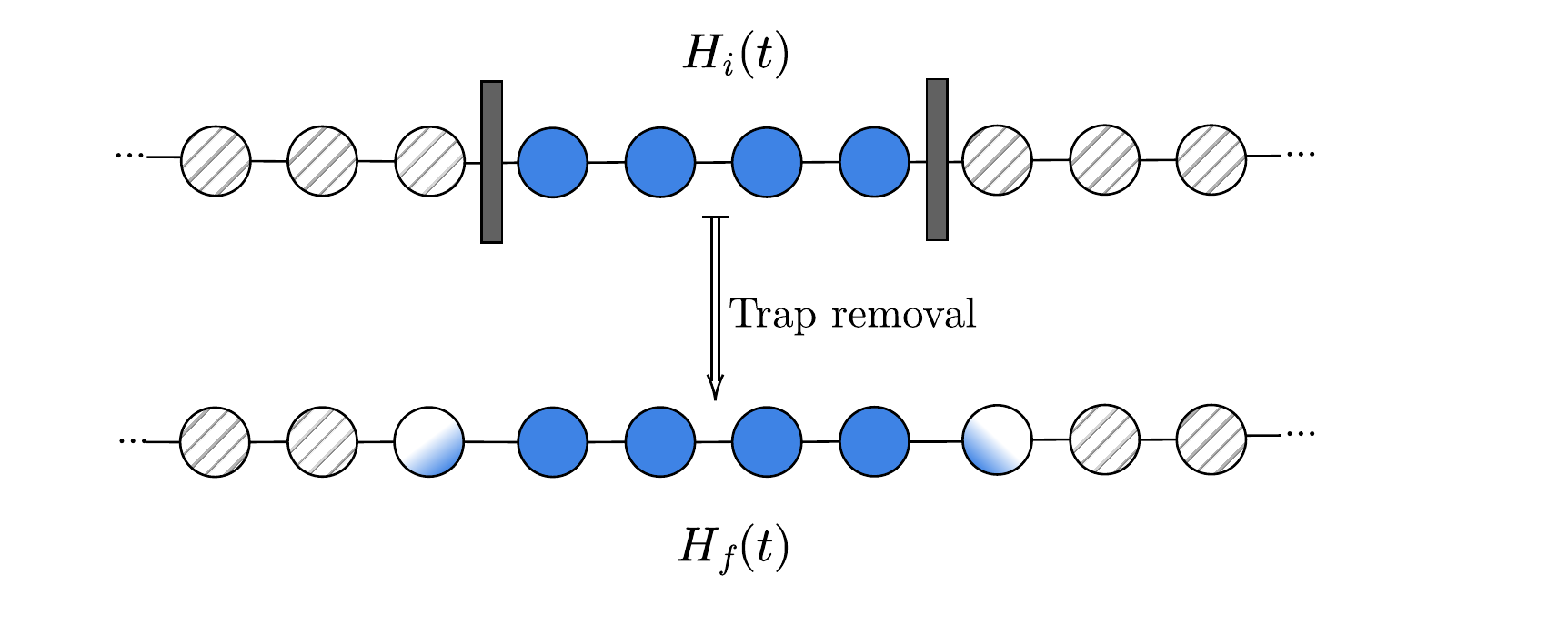}
\end{center}
        \caption{Schematic form of the expansion process. The particles are initially
confined to a small number of sites, and the system is prepared in the ground state of
the initial Hamiltonian $H_i$ (upper figure). Abruptly at $t=0$, the confining potential
is removed, allowing the particles to expand into the remainder of the lattice (lower figure). 
The Hamiltonian under which the system evolves during this expansion, $H_f$, may be the same or
different from $H_i$.}
        \label{schematic}
\end{figure}

In \Fref{schematic} we give a schematic representation of the process we simulate. Initially
the bosons are confined to a small part of the lattice (4 sites) and the system is prepared in its ground-state. The potential confining them to this small region is then instantaneously 
removed, allowing the cloud of particles to expand into the unoccupied portion of the lattice.
During this expansion we measure the density of the particle cloud, together with various
other one and two particle correlation functions. 
 
If the beginning of the expansion is made to coincide with the suppression or a strong reduction of the interactions (through the use of Feshbach resonances), TOF experiments can provide accurate information on the momentum distribution and correlations of the boson system before the expansion. This approach has been realized experimentally in Ref. \cite{13R0}.

\begin{figure}[htb]
        \centering
        \includegraphics[width=.9\textwidth]{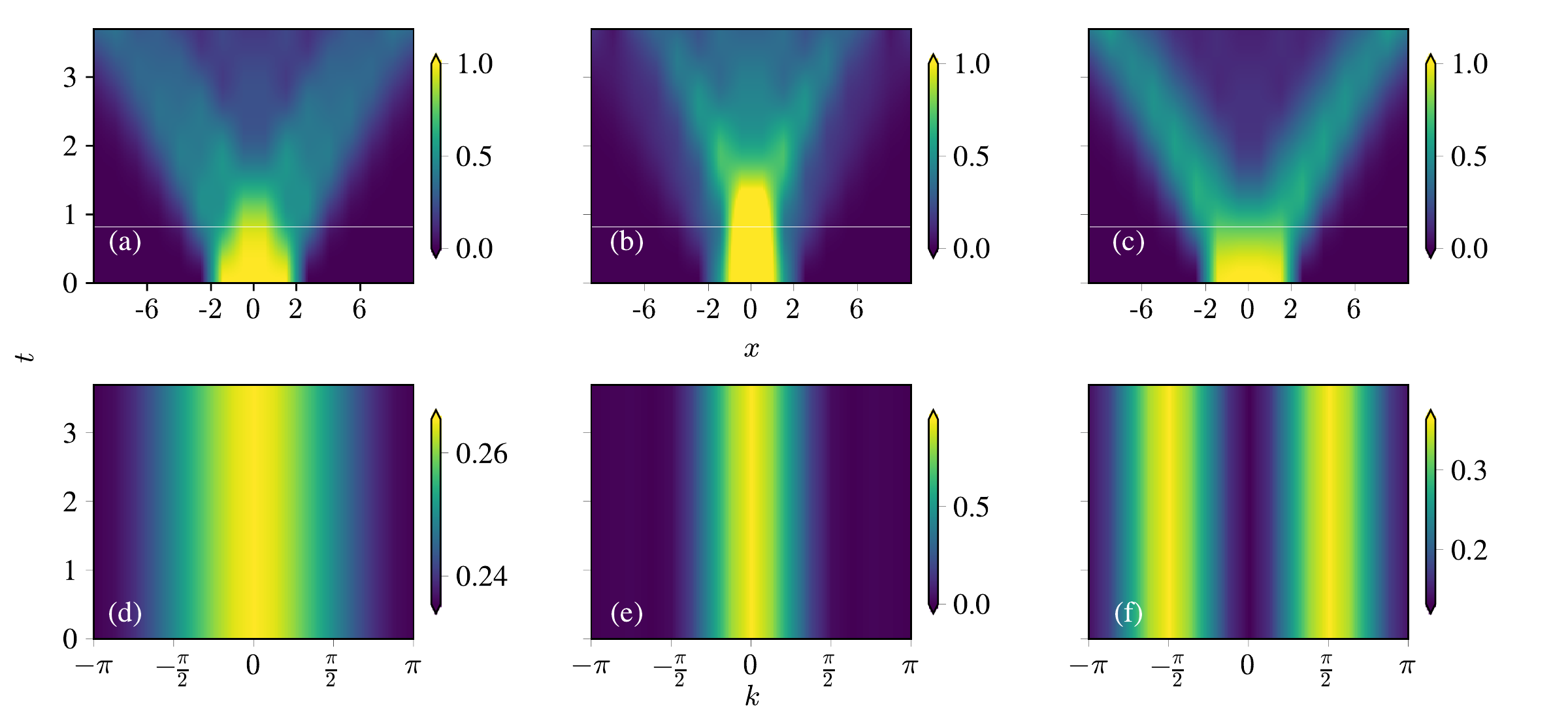}
        \caption{Time evolution of the densities in the case of interaction-free expansion, i.e., $U_f = 0$. Panels (a)-(c) correspond to $\rho(x,t)$ while (d)-(f) correspond to $\rho(k,t)$. Each column is assigned to one different initial state. Panels (a) and (d) show the densities for a MI (with $U_i=100$), while (b) and (e) correspond to a conventional superfluid ($U_i=0.1$).
                The cat state (obtained with driving parameter $\kappa_i = 0.5$) is shown in panels
                (c) and (f). Note the fine color resolution in panel (d), which would look uniform on the color scale of panels (e) or (f).}
        \label{fig:S3-rho-xkt}
\end{figure}

\Fref{fig:S3-rho-xkt} shows the results for the density in real and momentum space as a function of time under the assumption that $U_f=0$, for an expansion of 4 particles from 4 to 16 sites. The real-space density $\rho(x,t)$ is shown in the upper row. The absence of interactions during the expansion is reflected in the time constancy of $\rho(k)$ in all the cases considered. The expansion is tracked until times when the two wave packets are reflected at the hard walls or meet on the opposite side of the ring, the physics of both cases being very similar. For convenience we will focus on the case of a ring.

From left to right, the results correspond to the initial state being prepared in a conventional Mott insulating (MI) state (\Fref{fig:S3-rho-xkt}a,d), the conventional superfluid ground state (\Fref{fig:S3-rho-xkt}b,e), and the cat-state superfluid obtained from KD with a large driving amplitude (\Fref{fig:S3-rho-xkt}c,f). 

The first, leftmost column shows $\rho(x,t)$ and $\rho(k)$ for the expansion of a system initially prepared in the Mott insulating state of the conventional, undriven BH model. For $t<0$, $U_i=100$ while for $t>0$, $U_f=0$. As expected for the Mott insulator, the momentum distribution is essentially uniform, with a slight preference for small momenta around $k=0$ (note the fine color scale in \Fref{fig:S3-rho-xkt}d). The expansion in real space reflects a quick depletion of the initial, smaller chain due to the population of many states with nonzero momentum which fly away swiftly during the expansion.

The second column (\Fref{fig:S3-rho-xkt}b,e) shows the expansion of the boson gas initially prepared in the superfluid ground state of the conventional BH model with $U_i=0.1$. The system expands slowly in real space because it is initially condensed in $k=0$. The ability to expand relies on the population of momenta $k\neq 0$, which has two origins. One is the presence of a depletion cloud due to the initial interactions. Another reason is that the condensation at $k=0$ occurs in the space of the momenta that are allowed in a small chain of 4 sites. The extension to $L=16$ creates new allowed momenta, so that the momentum $k=0$ in the shorter chain mixes with several nonzero momenta of the longer chain. 

This inequivalence between the initial and final zero momentum states also plays a role in the expansion of a MI previously discussed. In both cases, it accounts for the slight reduction of the density in the central region, where zero momentum atoms are expected to remain.

We have also computed the expansion of a system initially in the Mott insulating state induced by kinetic driving with low amplitude $\kappa=0.1$. Except for a mild preference for the occupation of momenta $k=\pm \pi/2$ (instead of $k=0$ as in the conventional BH model, see \Fref{fig:S3-rho-xkt}d), the Mott state and its subsequent evolution is identical to the conventional MI. Since this pattern of essentially identical behavior of the two MI states is quite general, here and in future sections we only present results for the conventional Mott insulator.

The third column (\Fref{fig:S3-rho-xkt}c,f) shows the expansion of the boson gas initially prepared in the cat-like ground state of the KD Hamiltonian \eref{Heff}. The origin and structure of this exotic ground state has been analyzed in Refs. \cite{18PI,19PI,21MA}. The collective occupation of the momenta $\pm \pi/2$ translates into a fast and sharp expansion of the fragmented condensate in opposite directions (note that the momentum $\pi/2$ has the largest group velocity).

In addition to the lack of coincidence between the sets of initial and final available momentum states which we have already mentioned, there is another reason why, in the interaction-free expansion, the long-time real-space density may fail to faithfully reproduce the initial momentum distribution within the discrete and finite system we are considering. Unlike in free space, particles with momentum magnitude above $\pi/2$ move not faster but more slowly in a tight-binding lattice. This has little effect on the expansion of the conventional superfluid and cat-like ground states, but is quite relevant at long times in the case of an initial Mott insulator (not shown), where the initial population of those momenta is important.

\begin{figure}[h!]
        \centering
        \includegraphics[width=.9\textwidth]{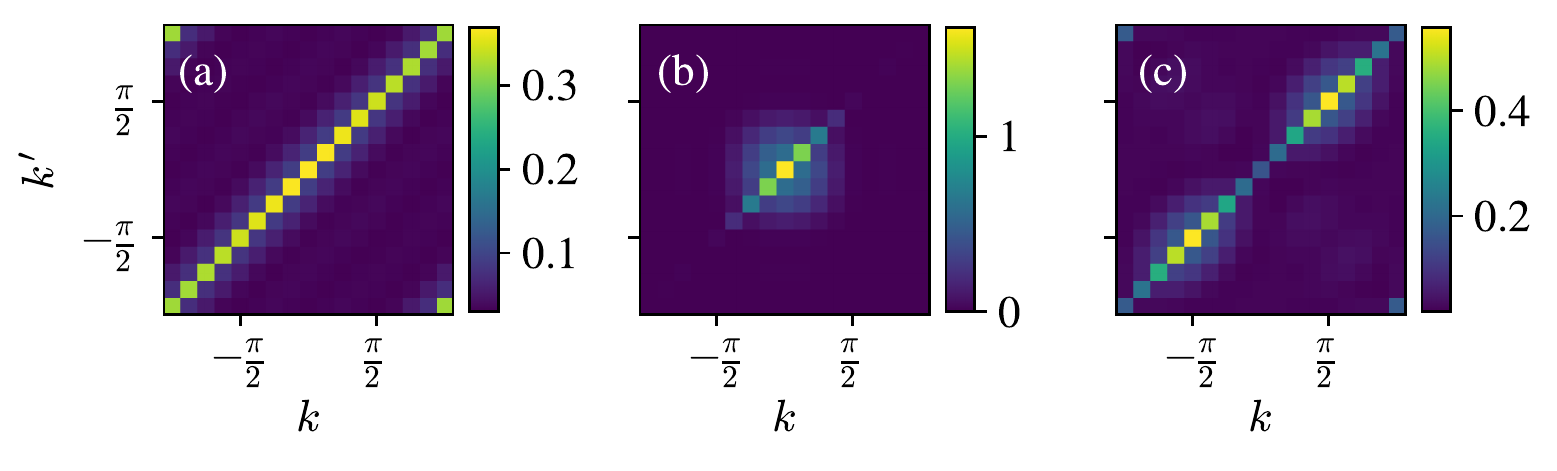}
        \caption{Two-particle density matrix $\rho(k,k')$ in the $U_f = 0$ case, measured at a positive time. Panels (a), (b), and (c) correspond to the initial states conventional MI, superfluid, and cat state, respectively. The choice of the particular time $t>0$ is irrelevant, because no forces operate during the expansion.}
        \label{fig:S3-rho-kk-p}
\end{figure}

\Fref{fig:S3-rho-kk-p} deals with the correlations in the occupation of the various momentum states. Due to the absence of interactions, the figures are the same for all times during the expansion. The conventional superfluid (\ref{fig:S3-rho-kk-p}b) shows a highly correlated occupation of small momenta near $k=0$. The case of the cat-like ground state of the kinetically driven system (\Fref{fig:S3-rho-kk-p}c) reveals a bimodal occupation of momentum states that is an extreme form of statistical correlation. Although to varying degrees, the correlated occupation of momentum states seems to be a universal feature of confined boson systems. Some exact results can be obtained in the limit of a MI.

\subsection{Momentum correlations in the Mott insulator.}
\label{mc-MI}

For the MI (\Fref{fig:S3-rho-kk-p}a), we observe that all momenta are occupied with essentially the same probability, as is expected from localized particles. The bosonic nature of the atoms is reflected in the highly correlated character of this uniform momentum occupation. This result can be understood if one notes that, for the MI state, the two-particle momentum density for $N$ bosons in a chain of $L$ sites (with $n=N/L$ an integer) is
\begin{equation}
	\langle n_k n_{k'}\rangle = n^2-n(n+1)/L+n(n+1)\delta_{kk'} \, ,
	\label{nknkp}
\end{equation}
which reveals a clear statistical preference for the correlated occupation of momentum states \footnote{The result \eref{nknkp} was given in Ref.  \cite{18PI} for the case $n=1$. Note also the different convention in the normalization of the momentum density.}. 

If we define the operators
\begin{equation}
N_+ = \sum_{k>0} n_k \, , \;\; N_- = \sum_{k<0} n_k \, ,
	\label{Npm}
\end{equation}
with $N_+ + N_-=N$ (for simplicity, we assume that $k=0$ does not exist or has a negligible weight), we obtain
\begin{equation}
	\langle N_+ N_\pm \rangle = [N^2 \pm (n+1)N]/4 \, .
	\label{NpNpm}
\end{equation}
The variance of the difference is
\begin{equation}
	v \equiv \left\langle \left(N_+ - N_- \right)^2 \right\rangle = (n+1)N \, .
	\label{variance}
\end{equation}
The large $N$ limit includes two interesting cases. For $n$ of order unity, $v \sim N$, while for $L$ of order unity, $v \sim N^2$. 

In our case, before the expansion we have $n=1$ and $N=4$. Equation \eref{nknkp} explains well the bunching in momentum space which we observe numerically. We have explicitly checked it for the case (4,4).

\section{Dynamic expansion\label{sec_dynamic}}

\begin{figure}[h!]
        \centering
        \includegraphics[width=.9\textwidth]{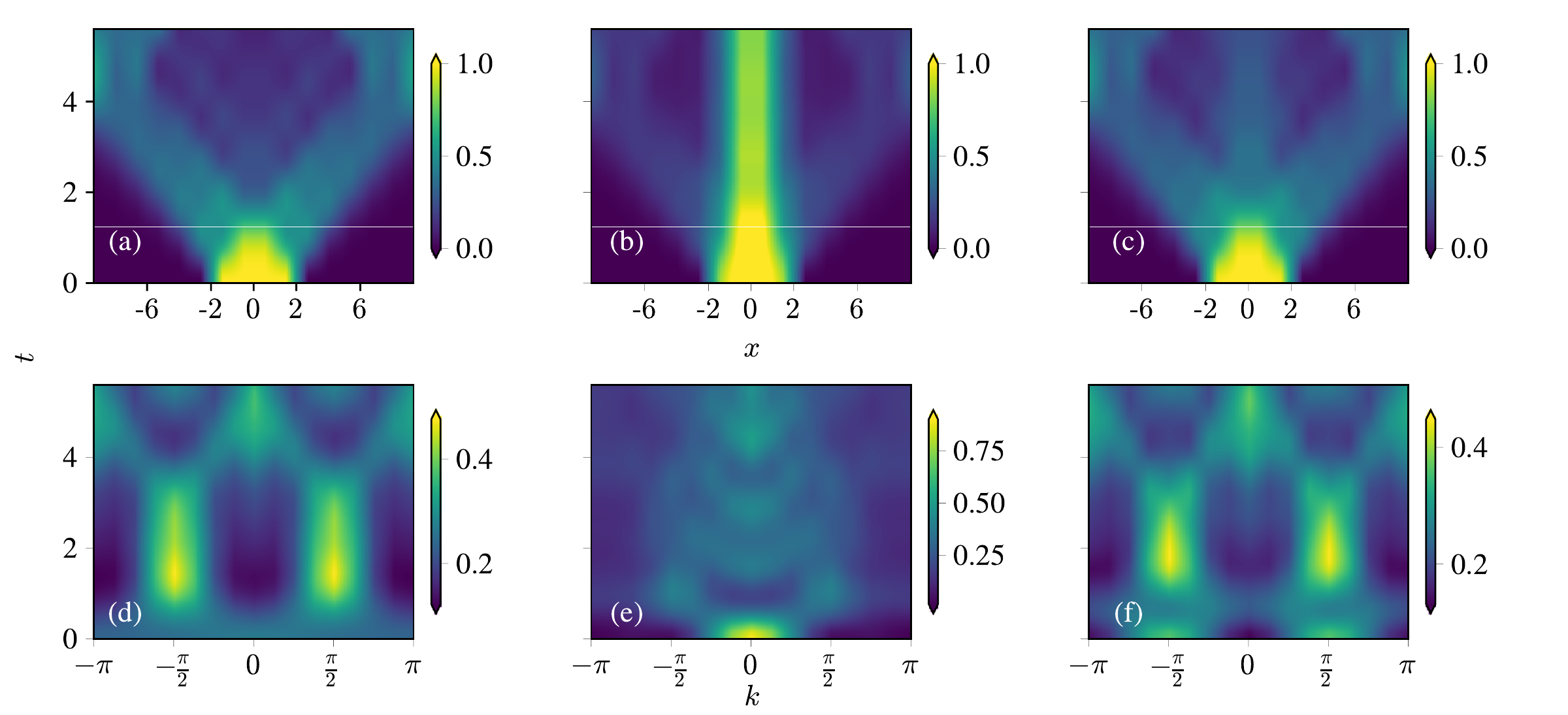}
        \caption{Time evolution of the densities in the case of a dynamic expansion, with $U_f= 100$. Panels (a)-(c) correspond to $\rho(x,t)$ while (d)-(f) correspond to $\rho(k,t)$.
                Each column is assigned to one different initial state.
                Panels (a) and (d) show the densities for a MI (with $U_i=100$), while (b) and (e) correspond to a conventional superfluid ($U_i=0.1$). The cat state (obtained with driving parameter $\kappa_i = 0.5$) is shown in panels (c) and (f).}
        \label{fig:S4-rho-xkt}
\end{figure}

In this section we study the expansion of the boson gas in the presence of strong interactions ($U_f=100$). In \Fref{fig:S4-rho-xkt} we show the real and momentum-space densities as a function of time for the cases where the boson system is initially prepared in a conventional Mott insulator (\Fref{fig:S4-rho-xkt}a,d), a conventional superfluid (\Fref{fig:S4-rho-xkt}b,e), and a cat-like superfluid obtained from kinetic driving (\Fref{fig:S4-rho-xkt}c,f). 
As in section \ref{sec_spectroscopy}, the time interval plotted has been chosen long enough to show the effect of the expanding many-body wave components meeting at the opposite side of the ring.

In \Fref{fig:S4-rho-xkt}a,d we see how the conventional MI quickly condenses into momenta $\pm \pi/2$. This reproduces the quasi-condensation predicted in Refs. \cite{04RI,05RI}, and experimentally confirmed in Ref. \cite{15VI} for the expansion of a Mott insulator. We find here that this dynamical condensation also occurs starting from other many-body states, but not in an identical manner.

Figure \ref{fig:S4-rho-xkt}c,f shows the evolution of the one-particle densities of the system initially prepared in the cat-like ground state resulting from kinetic driving with $U_i=1$ (although we know that its equilibrium configuration is independent of $U_i$). On a very short time scale, the momentum distribution quickly blurs into a relatively uniform distribution, but soon develops a quasi-condensation into momenta $\pm \pi/2$ very similar to that appearing when the initial state is a Mott insulator.

\Fref{fig:S4-rho-xkt}b,e shows the time-dependent densities in the expansion of an initially conventional superfluid. The system quickly evolves from a $k=0$ condensate to a fragmented condensate performing oscillations between the two types of condensates. This behavior may be ascribed to the greater coherence which results from having started with a condensate in a single momentum state. By contrast, the left and right columns of \Fref{fig:S4-rho-xkt} suggest that the involvement of a broad distribution of momenta gives rise to a longer lasting condensation into momenta $\pm \pi/2$. 

\begin{figure}[h!]
        \centering
        \includegraphics[width=0.9\textwidth]{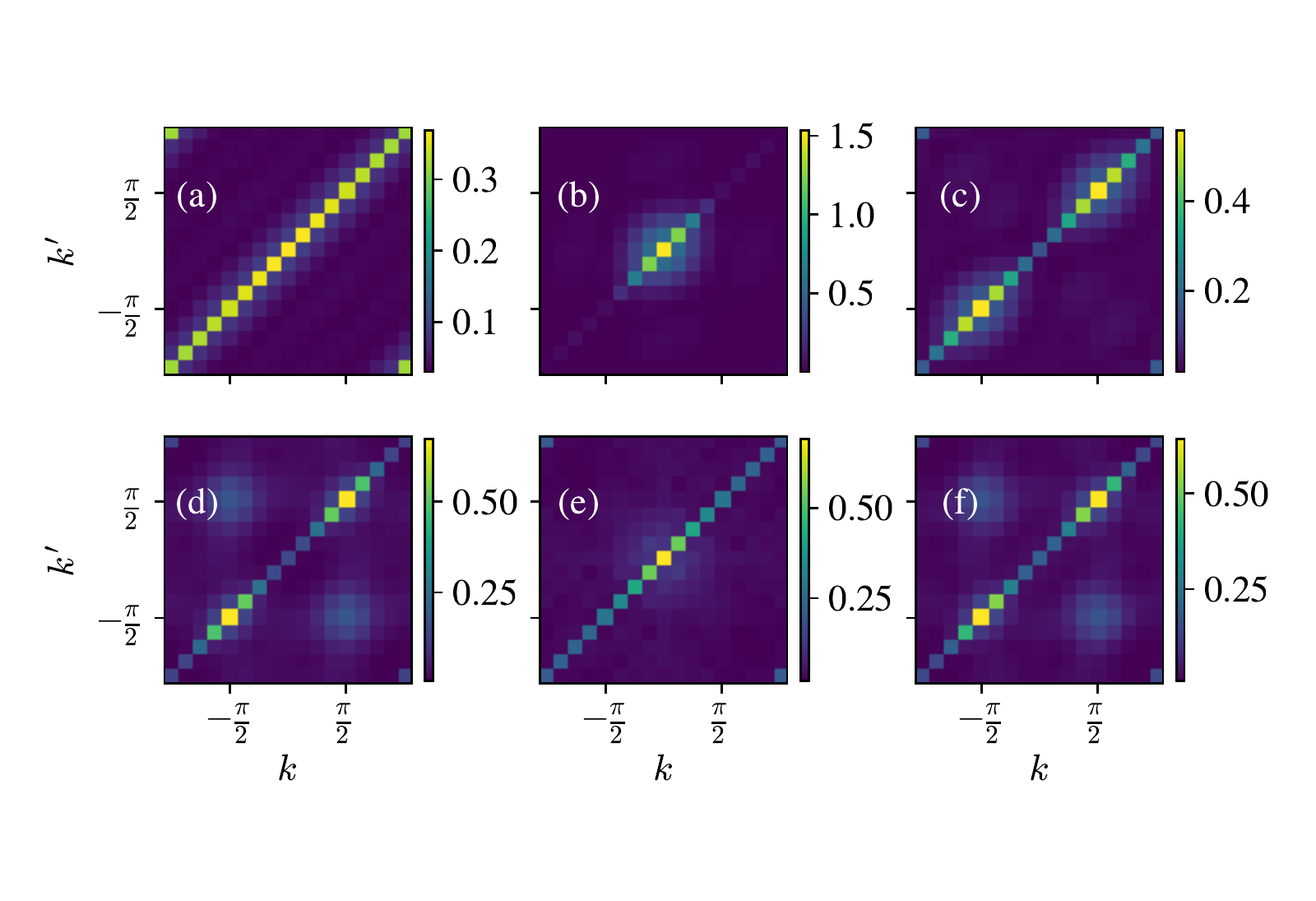}
        \caption{Two-particle density matrix $\rho(k,k')$ in the $U_f = 100$ case. Upper row: measurements at the initial time $t=0$. Lower row: measurements at time $t=1.4$. Panels (a) and (d) correspond to the MI, while panels (b) and (e) show the same quantity for the conventional superfluid. Lastly, panels (c) and (f) correspond to the initial cat state.}
        \label{fig:S4-rho-kk-p}
\end{figure}

\subsection{Momentum correlations in dynamically expanding quasi-condensates}
\label{mc-MI-dyn}

The studies of Ref. \cite{05RI,04RI} indicate that, starting from the MI, the quasi-condensation into momenta $\pm \pi/2$ is irreversible if the system is large enough and the expansion is sufficiently long lasting. We attribute the shorter lifetime of our nonzero momentum condensates to the specific geometry, as explained below.

A physical explanation of the dynamic condensation into momenta $\pm \pi/2$ is the following: As the cloud expands starting from a Mott insulator, atoms with different sign of the momentum move in opposite directions and soon stop overlapping spatially. At intermediate times, atoms only collide with other atoms that move in the same direction. The momentum distribution is basically uniform between 0 and $\pi$ for atoms moving to the right. In the pair collisions, the momentum must be conserved. The most likely value for the total momentum of an atom pair is $\pi$. Such pairs undergo collisions of the type
\begin{equation}
	\left(\frac{\pi}{2}+p,\frac{\pi}{2}-p \right)\longrightarrow \left(\frac{\pi}{2}+p',\frac{\pi}{2}-p' \right) \, .
	\label{pair}
\end{equation}
The case of $p'=0$ involves the double occupation of momentum $\pi/2$ and is thus enhanced due to bosonic amplification; the effective matrix elements for $p'=0$ are larger than for $p'\neq 0$.

This process tends to reinforce itself, as a higher occupation of $\pm \pi/2$ further 
favors the occupation of those momenta and, as a result, the occupation of $\pm \pi/2$ 
increases. Two pieces of data tend to support this explanation. 
The quasi-condensation into $\pm \pi/2$ occurs at the time when one expects the clouds 
of atoms with opposite velocities to fail to overlap spatially, considering that the 
maximum group velocity is 2. Another datum is the behavior of the nonzero momentum branches
when they encounter each other by wrapping around the ring geometry. 
Then one expects the most likely pair momentum to be zero, and so a condensate should 
begin to form at momentum zero. This is indeed what we observe at long times 
(see \Fref{fig:S4-rho-xkt}a,d).

Figure \ref{fig:S4-rho-kk-p} reveals the correlations in the system through the two-particle momentum density. As in the previous section (time-of-flight expansion without interactions), we find that the occupation of the various momentum states is fairly correlated. Comparing \Fref{fig:S4-rho-kk-p}c and \ref{fig:S4-rho-kk-p}f, we note that in the presence of interactions cat correlations resist but become less ideal.

Comparison of \Fref{fig:S4-rho-kk-p}d and \ref{fig:S4-rho-kk-p}f shows the remarkable fact that the initial MI insulator evolves towards a fragmented condensate with strong momentum correlations, almost identical to those remaining from the cat-like ground state of the initial KD system.

Whether the evolution toward cat-like momentum correlations starting from an initial MI will survive in the thermodynamic limit (large $N$ and $L$ with $n=N/L$ constant) seems an open question. On the one hand, the resemblance to the evolved KD state, with initial cat properties, suggests that cat correlations will survive. However, \eref{variance} reveals that the variance $v$ will grow with $N \sim L$, which is incompatible with cat-like correlations. But one might still argue that the conversion of the initial MI into two quasi-condensates of momenta $\pm \pi/2$ may be so fast that the effective number of available momenta for the expansion is not $L\sim N \gg 1$ but rather $L \sim 2$ with $N$ still $\gg 1$. In such a scenario, the variance will be of order $N^2$, which in this context can be considered a signature of cat-like correlations.  Whether the multi-channel or the two-channel picture will prevail in the thermodynamic limit is a question that merits further study.

\section{Expansion with kinetic driving and interactions \label{sec_kd}}

For completeness, we study in this section the expansion in the presence of kinetic driving. The parameters of the KD for $t>0$ are the same as those used in the previous sections for the KD ground state at $t<0$. As noted in Ref. \cite{18PI}, the ground-state properties of Hamiltonian \eref{Heff} are independent of $U$, which only sets a global time and energy scale. But in the dynamic process of expanding the KD system, or applying KD to a system not initially prepared with KD, the value of $U_f$ does matter. 

\begin{figure}[h!]
        \centering
        \includegraphics[width=.9\textwidth]{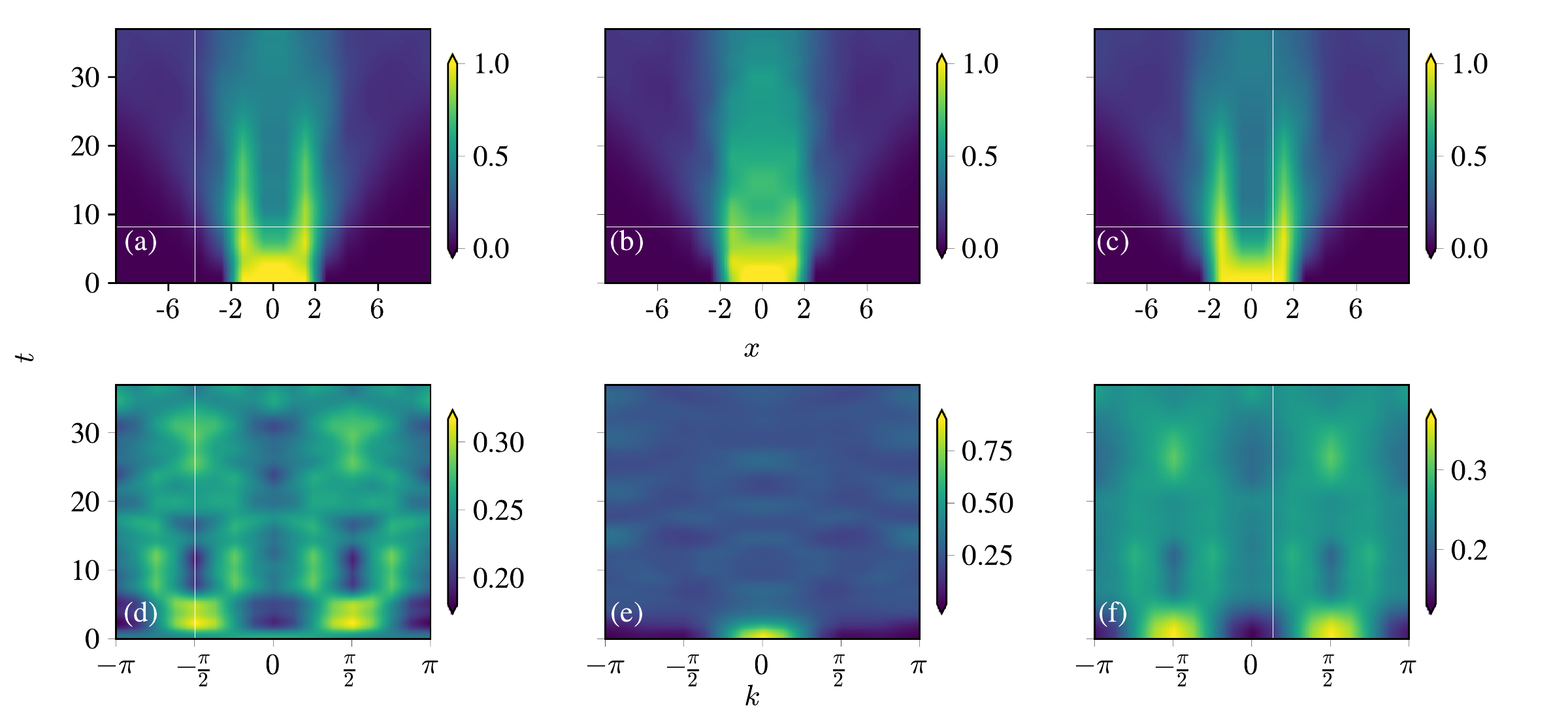}
        \caption{Time evolution of the densities in the case of a kinetically-driven dynamic expansion. Panels (a)-(c) correspond to $\rho(x,t)$, while (d)-(f) correspond to $\rho(k,t)$. Each column is assigned to one different initial state. Panels (a) and (d) show the densities for a MI (with $U_i=100$), while (b) and (e) correspond to a conventional superfluid ($U_i=0.1$).
                The cat state (obtained with driving parameter $\kappa_i = 0.5$) is shown in panels (c) and (f).}
        \label{fig:S5-rho-xkt}
\end{figure}

The results are shown in \Fref{fig:S5-rho-xkt}. Two physical trends are noteworthy. The expansion is very slow, as reflected in the modest increase of the central cloud on a time scale substantially longer than in previous sections. This is expected for a system with zero average kinetic energy. In fact, in Ref. \cite{19PI} it was shown that the (exotic) current operator has a zero expectation value in both branches of the cat-like ground state. The difference between the two branches lies in the momentum distribution, which translates into opposite-moving wave packets upon expansion without KD, as we saw in section \ref{sec_spectroscopy}.

Naively, one might expect the KD prepared ground state to remain localized during the expansion with KD applied, as its current expectation value should remain zero. However, the current that vanishes is that defined in the shorter chain of the initial state. Once the chain is expanded, from $L=4$ to $L=16$, the current operator changes and the branches of the initial cat-like ground state no longer have a vanishing expectation value of the current operator. As a consequence the system can expand, but only slowly.

A second physical feature worth noting has to do with the absence of single-particle hopping. As emphasized in Refs. \cite{18PI,19PI}, only even-order hopping process involving two particles are permitted. This means that for a particle to jump two sites, it needs the assistance of a nearby particle which may also jump or just remain idle. This means that as the expansion starts, particles within the bulk of the initial cloud can move faster towards the boundaries because they can be assisted by other particles. However, once they reach the boundaries, hopping is still possible but is more difficult due to the scarcity of nearby assisting particles. This results in a depletion of the central region of the cloud and an accumulation of particles at the initial boundaries, as can be clearly seen in \Fref{fig:S5-rho-xkt}a,c. This accumulation effect is less marked in \Fref{fig:S5-rho-xkt}b because of the initial large population of the $k=0$ state, which results in a slower depletion of the initial central region.

It is interesting to compare the evolution of initial states as different as the MI and the KD superfluid (\Fref{fig:S5-rho-xkt}a,c). The depletion of the central region is slightly faster for the cat-like state, as it is mostly populated by particles with a large group velocity. But at longer times, the evolution is very similar, especially in the position representation.

\section{Speed of expansion\label{sec_velocity}}

We may study the velocity scales involved in the problem by focusing on the time evolution of the standard deviation of the position distribution $R(t)$ defined as
\begin{equation}
	R(t) = \left[\frac{1}{N}\sum_{x}  x^2 \rho(x,t)\right]^{\frac{1}{2}} \, ,
\end{equation}
where we exploit the space inversion symmetry of the problem around $x=0$. Then, following Ref. \cite{13R0}, we may define the radial velocity
\begin{equation}
	\label{speed}
	v_r(t) =\dot{R}(t)\, .
\end{equation}

Another relevant velocity is the standard deviation of the group velocity before the expansion
\begin{equation}
	v_g = \left[\frac{1}{N}\sum_{k}(\sin k)^2 \rho(k)\right]^{\frac{1}{2}} \, ,
	\label{vg}
\end{equation}
which, in a spectroscopic TOF, remains time independent.  

\begin{figure}[h!]
        \centering
        \includegraphics[width=.75\textwidth]{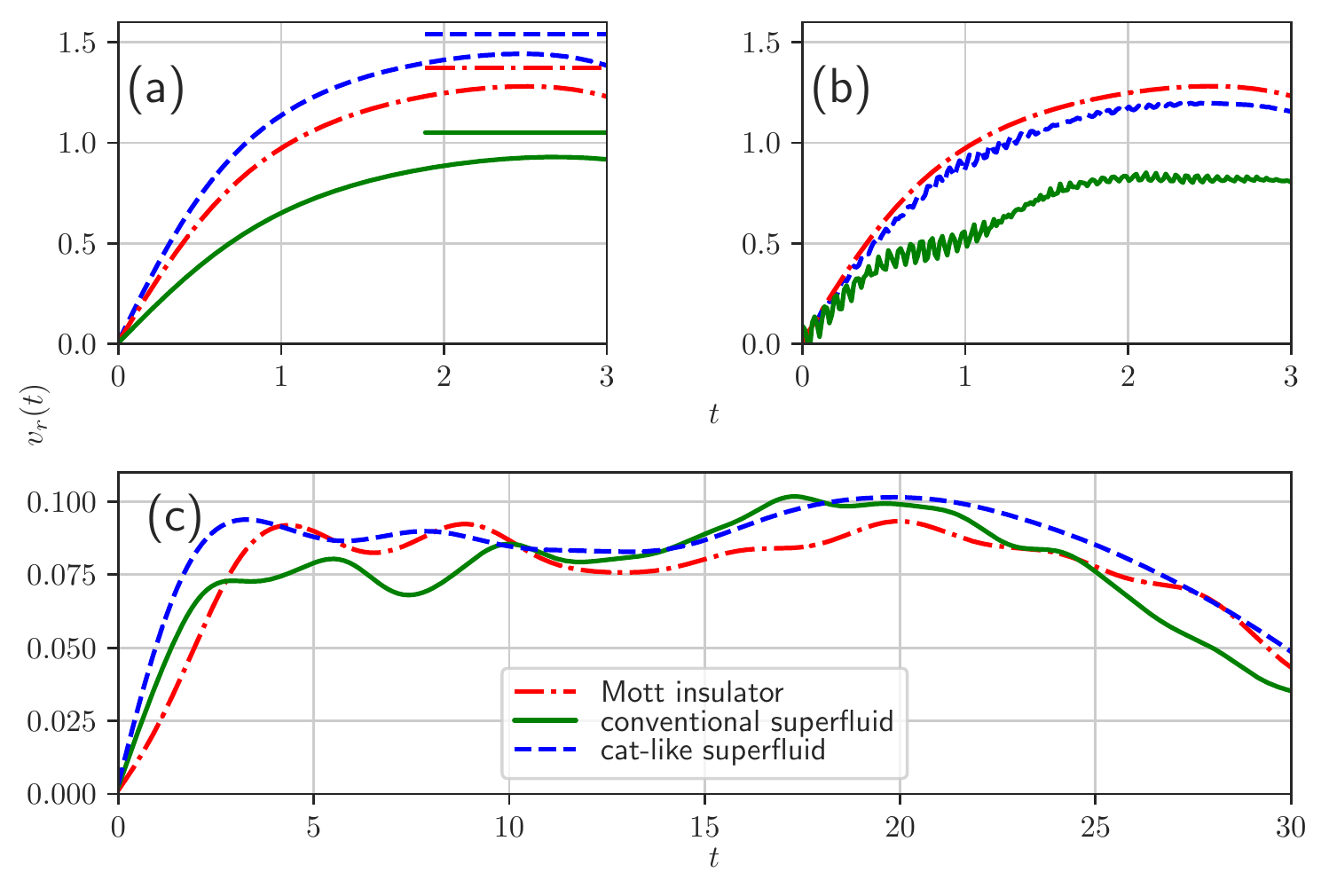}
        \caption{Radial velocity $v_r$ for the cases considered in each section. (a) Spectroscopic TOF (section \ref{sec_spectroscopy}), (b) dynamic expansion (section \ref{sec_dynamic}), (c) expansion under kinetic-energy driving (section \ref{sec_kd}).  Horizontal lines in (a) represent the standard deviation of the group velocity at the start of the expansion [Eq. \ref{vg}].}
        \label{fig:speed}
\end{figure}

In \Fref{fig:speed}a we show the evolution of the radial velocity during an expansion with $U_f$=0. We also show, in horizontal segments, the average group velocity as defined in \eref{vg} evaluated at $t=0$. At long times, in an unlimited expansion, we would expect $v_r(t)\rightarrow v_g$. However, we do not observe this because the expanding wave packets meet at the opposite side of the ring. For that reason, in the three initial cases considered here (MI, conventional superfluid, or cat-like superfluid), $v_r(t)$ systematically falls short of reaching $v_g$.

Of those three cases, the cat-like superfluid reaches the highest velocity because it starts with a large population of particles in momenta of magnitude $\pi/2$, which has the largest group velocity in the lattice. The slowest expansion is performed by the initial conventional superfluid, which predominantly occupies the zero momentum state. The MI state, with a uniform distribution over all momenta in the Brillouin zone represents an intermediate case.

\Fref{fig:speed}b shows the evolution of the radial velocity during a dynamic expansion with a large interaction, as in section \ref{sec_dynamic}. The similar evolution of the three initial systems at short times, and of the MI and cat state at all times, underlines the strong effect of interactions during the expansion.
On the other hand, the leveling off of $v_r(t)$ is due to the encounter of the two expanding wave packets on the opposite side of the ring, as discussed for \Fref{fig:speed}a.

In \Fref{fig:speed}c we show $v_r(t)$ in the presence of kinetic-energy driving. As expected, the velocity scales are much smaller and somewhat fluctuating. 
We notice that the decrease of the radial velocity at long times does not have a geometric origin (because the expanding branches have not yet met at the opposite side of the ring) but rather is due to the decrease in density, which tends to inhibit correlated hopping.

\section{Conclusions\label{sec_conclusions}}

We have made an exact numerical study of the expansion of a one-dimensional Bose gas under three different conditions and starting from three different initial configurations. The system considered has been that of 4 bosons in 4 sites suddenly allowed to symmetrically occupy 16 sites which, for convenience, we assumed to be arranged in a ring geometry. Despite the smallness of the system, we have found a surprisingly rich variety of physical properties in both the equilibrium and the non-equilibrium phases.

Although our initial motivation has been to investigate how a time-of-flight experiment can be used to probe the different ground states of a kinetically driven BH model, we have extended our study to the ground states of the conventional, undriven BH model. This allows comparison with a well understood system, while giving us the opportunity to gain some additional insight on the conventional case.

The initial configurations considered have been the Mott insulator, the superfluid state of the conventional BH model, and the cat-like ground state of the kinetically driven boson gas. 
Since the properties of the MI ground states of the conventional and the KD Bose gas are practically identical, we have discussed only the conventional MI.

In all cases we have considered the expansion under three different conditions. The most informative one is the spectroscopic TOF, with interactions tuned to zero, as it gives a detailed account of the momentum distribution and its correlations, except for high initial momenta. 

The dynamic expansion is also interesting because it is the most natural one. 
For the case of an initial Mott insulator, we have reproduced the dynamical quasi-condensation into momenta $\pm \pi/2$ and have given a physical explanation of its origin in terms of interacting bosons, which complements the original, more mathematical discussion of Ref. \cite{04RI}. We have pointed out that those quasi-condensates form with important, cat-like momentum correlations, and have argued that whether such correlations will survive in the thermodynamic limit is an open question.

We have also considered the case in which the kinetic-energy driving remains switched on during the expansion. Due to the suppression of the single-particle kinetic energy, the expansion under KD is slow, but we conjecture that for high densities it can be considerably faster due to the important role of assisted hopping.

We include an Appendix where we show that some popular measures of the two-particle momentum correlation tend to hide the possible existence of cat-like correlations by giving too much weight to lowly populated momentum states. 

This work should stimulate further research on the expansion of conventional boson systems. In particular, the nature of momentum correlations during the expansion remains a deep open question. As to the boson systems prepared under kinetic driving, the present work shows a preliminary vista of the richness of this novel form of quantum matter.

\ack
This work has been supported by Spain's MICINN through Grant No. 
FIS2017-84368-P and by Universidad Complutense de Madrid through 
Grant No. FEI-EU-19-12. One of us (FS) would like to acknowledge
the support of the Real Colegio Complutense at Harvard and the 
Harvard-MIT Center for Ultracold Atoms, where part of this work was done.

\appendix
\section{Two-particle momentum correlations \label{app_g2}}

In the literature (see e.g. Ref. \cite{19CA}), it is common to define the second-order correlation function
\begin{equation}
	G^{(2)}(k,k') = \langle a^{\dagger}_{k} a^{\dagger}_{k'} a_{k'}a_{k} \rangle \, ,
\end{equation}
or its normalized version,
\begin{equation}
	g^{(2)}(k,k') = \frac{\langle a^{\dagger}_{k} a^{\dagger}_{k'} a_{k'}a_{k} \rangle}{\rho(k)\rho(k')} \, ,
\end{equation}
where $\rho(k)=\langle n_k \rangle$ is the momentum density. The previous definition is connected to the two-particle momentum density $\rho(k,k')=\langle n_k n_{k'} \rangle$ through the relation
\begin{equation}
	G^{(2)}(k,k') 
	= \rho(k,k') - \rho(k)\delta_{kk'} \, .
	\label{G2k-r2k}
\end{equation}

\begin{figure}[h!]
	\centering
	\includegraphics[width=.9\textwidth]{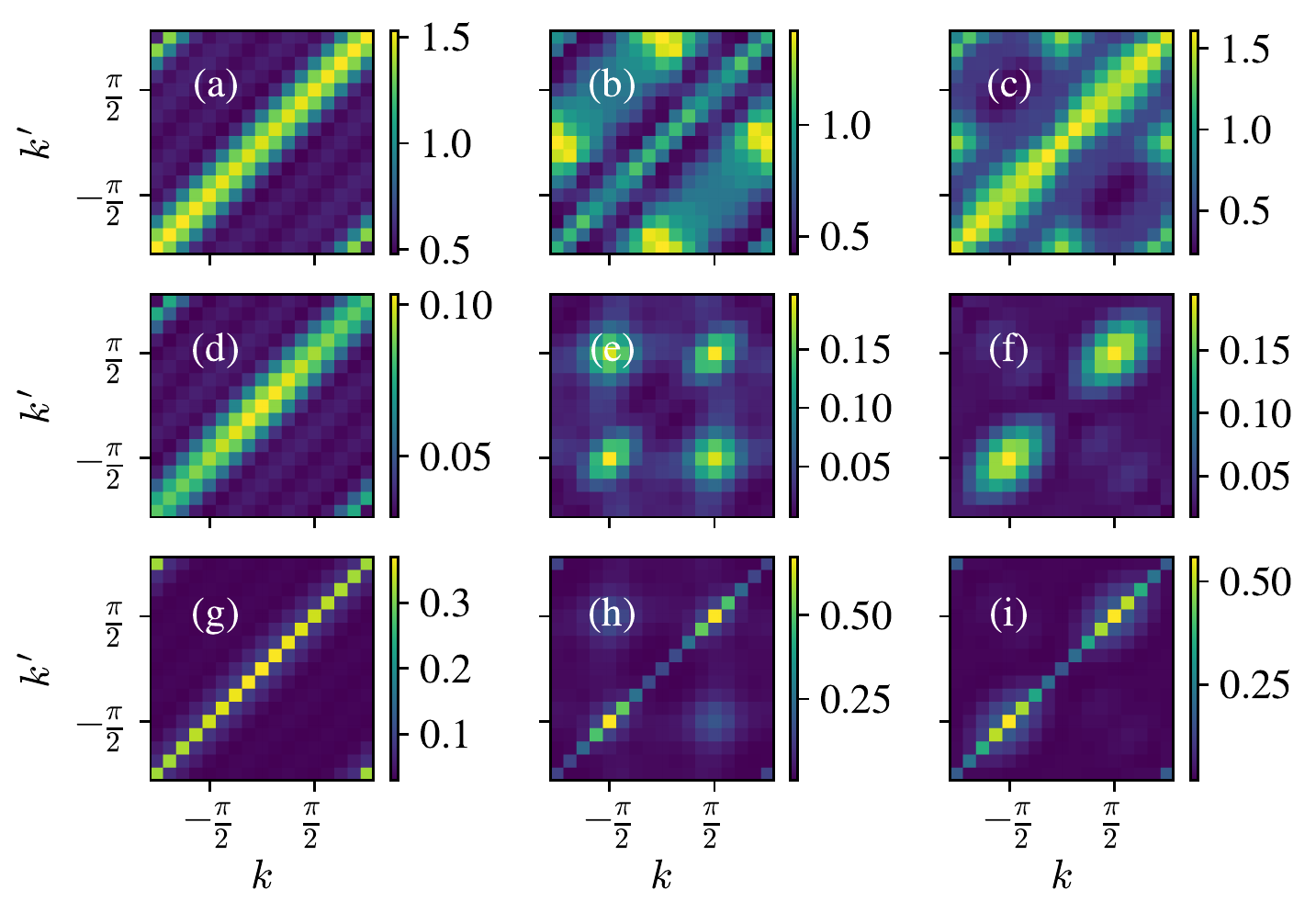}
	\caption{Comparison between $g^{(2)}(k,k')$, $G^{(2)}(k,k')$, and $\rho(k,k')$. The upper row shows $g^{(2)}(k,k')$, the middle one shows $G^{(2)}(k,k')$ and the lower one shows $\rho(k,k')$. The left column [panels (a), (d), (g)] corresponds to the initial Mott insulator 
	at time $t=0$. The central column [panels (b), (e), (h)] corresponds to the dynamical quasicondensate
	at time $t=1.4$. Lastly, panels (c), (f), (i) correspond to the cat state 
	at time $t=0$.}
	\label{fig:G2}
\end{figure}

In \Fref{fig:G2} we present these three measures of the two-particle momentum correlation, namely, $g^{(2)}(k,k')$, $G^{(2)}(k,k')$, and $\rho(k,k')$. We can see that the cat state is manifest in \Fref{fig:G2}f or \Fref{fig:G2}i but not in \Fref{fig:G2}c. If we focus only on the diagonal term, we can appreciate that $g^{(2)}(k,k)$ enhances the region around $k=0$ where the momentum density is low. The regions with a macroscopic occupation ($k=\pm\pi/2$) are then eclipsed. This is the reason why we favor the use of $\rho(k,k')$ over $g^{(2)}(k,k')$. With the use of the latter correlation function, cat states, or some many-body states with a large momentum variance, may pass unnoticed. 

For comparison we have also included the three correlation measures
for a conventional Mott insulator (\Fref{fig:G2}a,d,g) and for the dynamical quasicondensate (\Fref{fig:G2}b,e,h) at time $t=1.4$.


The broader peaks in $G^{(2)}(k,k')$, as opposed to those of $\rho(k,k')$, can be ascribed to the reduced difference between the cases $k=k'$ and $k\neq k'$ [see Eqs. \eref{nknkp}, \eref{G2k-r2k}], which tends to flatten the peaks. 

Finally, 
we note that \Fref{fig:G2}g, \Fref{fig:G2}h, \Fref{fig:G2}i contain the same information as 
\Fref{fig:S3-rho-kk-p}a,  \Fref{fig:S4-rho-kk-p}d, \Fref{fig:S3-rho-kk-p}c, respectively.

\newpage
\section*{References}
\bibliographystyle{iopart-num}
\bibliography{tof_bib}

\end{document}